\documentclass[sn-mathphys]{sn-jnl}% Math and Physical Sciences Reference Style
\usepackage{bm}

\jyear{2022}%

\begin{document}

\title{Reducing roundoff errors in numerical integration of planetary ephemeris}

\author{\fnm{Maksim} \sur{Subbotin}}

\author{\fnm{Alexander} \sur{Kodukov}}

\author*{\fnm{Dmitry} \sur{Pavlov}}\email{dapavlov@etu.ru}

\affil{\orgdiv{Faculty of Computer Science and Technology}, \orgname{St.~Petersburg Electrotechnical University}, \orgaddress{\street{ul. Professora Popova 5}, \city{St.~Petersburg}, \postcode{197022}, \country{Russian Federation}}}

\abstract{Modern lunar-planetary ephemerides are numerically integrated
  on the observational timespan of more than 100 years (with the last
  20 years having very precise astrometrical data). On such long
  timespans, not only finite difference approximation errors, but also
  the accumulating arithmetic roundoff errors become important because
  they exceed random errors of high-precision range observables of
  Moon, Mars, and Mercury. One way to tackle this problem is using
  extended-precision arithmetics available on x86 processors. Noting
  the drawbacks of this approach, we propose an alternative: using
  double-double arithmetics where appropriate. This will allow to use
  only double precision floating-point primitives which have
  ubiquitous support.}


\maketitle

\section{Introduction}\label{sec:intro}
A lunar-planetary ephemeris is essentially a table that contains
coordinates of Solar system planets and the Moon on a temporal
grid. Lunar-planetary ephemerides are used for various purposes,
including Earth and space navigation, prediction of orbits of
asteroids and spacecraft, prediction of solar and lunar eclipses, and
testing scientific hypotheses about fundamental properties of
spacetime. Since 1960s, ephemerides are built on computers with
numerical routines integrating equation of motion. Modern lunar-planetary ephemerides
(in order of first release) are: JPL DE \cite{DE440}, IAA EPM
\cite{EPM2021}, INPOP IMCCE \cite{INPOP21}. Also there exist
ephemerides based on analytical theories instead of numerical
integration, e. g. VSOP \cite{VSOP2013}, but they are outside the
scope of this work.

Equations of motion contain many free parameters that are determined
from astrometrical observations. Throughout decades, the accuracy of
ephemerides improved, following the improvement of the accuracy of the
observations and improvement of mathematical models.

As a rule, the accuracy of a mathematical model must be the same or
better than the accuracy of observations. In this work, we will
be focusing on the orbits of three bodies with the most precise
observations to date:
\begin{itemize}
\item The Moon --- most precise lunar laser ranging normal points
  reach 1 mm accuracy \cite{Murphy2012,Courde2017}.
\item Mars --- most precise spacecraft radio ranging normal points
  reach 55 cm postfit rms \cite{Kuchynka2012}.
\item Mercury --- MESSENGER spacecraft radio ranging normal points
  reach 60 cm postfit rms \cite{EPM2021}.
  \end{itemize}

The problem we are dealing with in this work is the arithmetical
roundoff errors that accumulate during the numerical integration of
equations of motion, therefore affecting the accuracy of ephemeris.

\section{Equations of motion}
In this work, we use the dynamical model of EPM2021 ephemeris, though
other ephemerides have similar dynamical models. The following bodies are
included into the model: 

\begin{itemize}
\item All planets;
\item The Sun and the Moon;
\item Pluto and 30 other transneptunian objects;
\item 277 asteroids;
\item Discrete uniform 180-point annulus approximating the asteroid belt;
\item Discrete uniform 160-point ring approximating the Kuiper belt.
\end{itemize}

For details of how the bodies were chosen, we refer to
\cite{Pitjeva2018,PitjevaKuiperBelt,KanPavlov}.
Sixteen bodies (the Sun, the planets, Pluto, Ceres, Pallas, Vesta, Iris, Bamberga)
obey Einstein--Infeld--Hoffmann equations of motion written up to $1/c^2$ terms:

\begin{equation}\label{eq:eih}
  \begin{aligned}
\bm{a}_i & = \sum_{j \neq i} \frac{Gm_j \bm{r}_{ij}}{r_{ij}^3} \\
& + \frac{1}{c^2} \sum_{j \neq i}
    \frac{Gm_j \bm{r}_{ij}}{r_{ij}^3}
    \left[  v_i^2+2v_j^2 - 4( \bm{v}_i \bm{v}_j) - \frac{3}{2} \frac{(\bm{r}_{ji} \bm{v}_j)^2}{r_{ji}^2} \right. \\
 & \qquad \qquad \qquad \qquad  \left. -\  4 \sum_{k \neq i} \frac{Gm_k}{r_{ik}} -
  \sum_{k \neq j} \frac{Gm_k}{r_{jk}} + \frac{1}{2}( \bm{r}_{ij}  \bm{a}_j )
  \right] \\
& + \frac{1}{c^2} \sum_{j \neq i} \frac{Gm_j}{r_{ij}^3}\left[\bm{r}_{ji}(4\bm{v}_i-3\bm{v}_j)\right](\bm{v}_i-\bm{v}_j) \\
& + \frac{7}{2c^2} \sum_{j \neq i}{ \frac{Gm_j \bm{a}_j }{r_{ij}}}
\end{aligned}
\end{equation}

\noindent where $\bm{a}$ is acceleration; $\bm{v}$ is velocity; $Gm$ is
the gravitational parameter; $\bm{r}_{ij} = \bm{r}_j -
\bm{r_i}$ is the position of body $j$ w.r.t. body $i$; and $c$ is
the speed of light.

Note that the first term in Eq.~(\ref{eq:eih}) is Newtonian
acceleration; also note that the accelerations are present also on the
right hand side of Eq.~(\ref{eq:eih}), making it look as an implicit
equation.  However, in practice, the $\bm{a}_j$ terms are replaced
with Newtonian accelerations, which has negligible effect on the
result.

Other point masses in the model interact with those 16 with only
Newtonian forces and do not interact with each other. This compromise
is made to speed up the calculations, with sub-centimeter impact on
the orbits of the planets and sub-millimeter impact on the
geocentric orbit of the Moon on the timespan of 40 years.

Other effects, important for dynamical model of ephemeris, were
omitted in this study. Those are: accelerations from solar oblateness
and Lense--Thirring effect; ``point mass--figure'' accelerations of the
Earth that come from the Sun, Venus, Mars, Jupiter, and the
Moon; the motion of the Moon as an elastic body with a rotating liquid
core \cite{Pavlov2016} experiencing tidal forces modeled with a delay
differential equation \cite{DDE}.

The equations are numerically integrated with the step of $1/16$ days
with an Adams--Bashforth--Moulton multistep predictor-corrector scheme
(PECEC mode)
of order 13. The implementation used in this work was analogous
to one described in \cite{DDE}, but without the ``delay'' part. The
formulas of predictor and corrector are given in
Eq.~(\ref{eq:pred-corr}).

\begin{equation}\label{eq:pred-corr}
  \begin{aligned}
    \bm{x}_{n+1}^{\mathrm{(p)}} &= \bm{x}_n +  h \sum_{j=0}^{k-1} \gamma_j \nabla^j \bm{f}_n \\
    \bm{x}_{n+1} &= \bm{x}_{n+1}^{\mathrm{(p)}} + h \gamma_k \nabla^k \bm{f}_{n+1},
  \end{aligned}
\end{equation}

\noindent Here, $\bm{x}_n$ is the state of the dynamical system, $\bm{f}_n
= \bm{f}(\bm{x}_n)$ is the time derivative of $\bm{x}$, $\nabla$ is
the backward finite difference operator, $h$ is the step size, $n$ is the
step number, $k$ is the order of the method, and $\gamma_j$ are
precalculated coefficients.

For obtaining the $\bm{x}_2 \ldots \bm{x}_{k}$, a single-step
Dormand--Prince scheme of order 8 is used (with the step size of
$h/8$).

The arithmetical roundoff errors happen in the numerical scheme itself
as well as in the calculation of $\bm{f}$.

\section{Estimation of numerical roundoff errors}
\subsection{Representation of numbers}
The most common numerical data type in computers is IEEE 754
\textit{double precision} --- a 64-bit binary floating-point data type
where the fractional part occupies 52 bits. Every number represented
in this format is assumed to have an error which does not exceed half
of the \textit{unit in the last place} (ulp), i.e. the value of the
last bit of the fractional part. For example, $1 + 2^{-52}$ is
represented exactly in double precision, while anything between 1 and
$1 + 2^{-53}$ is rounded down to $1$, and anything between $1 +
2^{-53}$ and $1 + 2^{-52}$ is rounded up to $1 + 2^{-52}$.  This gives
the rough estimate of double precision representation being accurate
to about $-\log_{10}2^{-53} \approx 16$ decimal digits, and the
relative error of representation of numbers hence being about
$10^{-16}$.

This is more than enough precision to represent astronomical data. For
example, the maximum Earth--Moon distance is approximately 400000 km,
while the best observations (see Sec. \ref{sec:intro}) have the
accuracy of about 1 mm; so at any point in time, it makes sense to
require the accuracy of the geocentric position of the Moon being 1 mm
at best\footnote{The semimajor axis of the orbit of the Moon in the
ephemeris strongly correlates with X-coordinates of retroreflectors;
hence, the said coordinates have the uncertainty of about
3 cm~\cite{Pavlov2019} at best. However, it still makes sense to require
better accuracy in lunar ephemeris when geocentric coordinates of
the retroreflectors are of interest; in those coordinates, the uncertainties
of the X-coordinates of retroreflectors and the semimajor axis of the orbit
of the Moon are largely canceled out.};
that is, the relative error of representation of $2.5 \times 10^{-12}$ will be enough.

Similarly, the maximum Earth--Mars distance is about 400 million km;
to match the accuracy of observations of about 0.5 m, the
relative error of $1.25\times 10^{-12}$ will be enough.

However, during the process of numerical integration,
each arithmetical operation involving double precision
numbers rounds its result to double precision again, and this roundoff
error propagates to further calculations; as a result, the final
orbits suffer from numerical errors that accumulate.

\subsection{Two-way integration}
In order to estimate the roundoff errors, one must find a way to
separate them from approximation errors inherent to the numerical
scheme and the errors in the underlying model. This is achieved via
integration of the same equations forward and then backward over time.
The backward integration uses the result of the forward integration as
initial state. Mathematically, the two integrations should produce
identical results. In reality, both integrations have model errors,
approximation errors and numerical errors---but the errors of the
first two kinds cancel out\footnote{In a multistep scheme, the
approximations are slightly different near the ends of the timespan of
integration due to the ``warm-up'' stage, but as we will later see,
this difference does not make any noticeable impact.}. The comparison
of the orbits obtained with the two integrations thus gives us the
estimate of the accumulated roundoff error. For previous usage of this
technique, see e.g.~\cite{INPOP06}.

In modern ephemeris, the initial parameters of orbits are determined
from observations that cover the interval of about 100 years. The
\textit{epoch}---the date those parameters are referred to---is placed
somewhere inside this interval, and the numerical integration is then
performed in both directions from epoch. To reduce the overall error
of the numerical integration, the epoch could be in the middle
of the interval; but to account for the better precision of
observations of modern times, the epoch is usually shifted right from
the middle. In EPM, the epoch is Oct 27, 1984 (JD 2446000.5). In this
work we will study the accumulation of roundoff error on the
timespan of 40 years of integration (in one direction).

\begin{figure}[p]
  \input{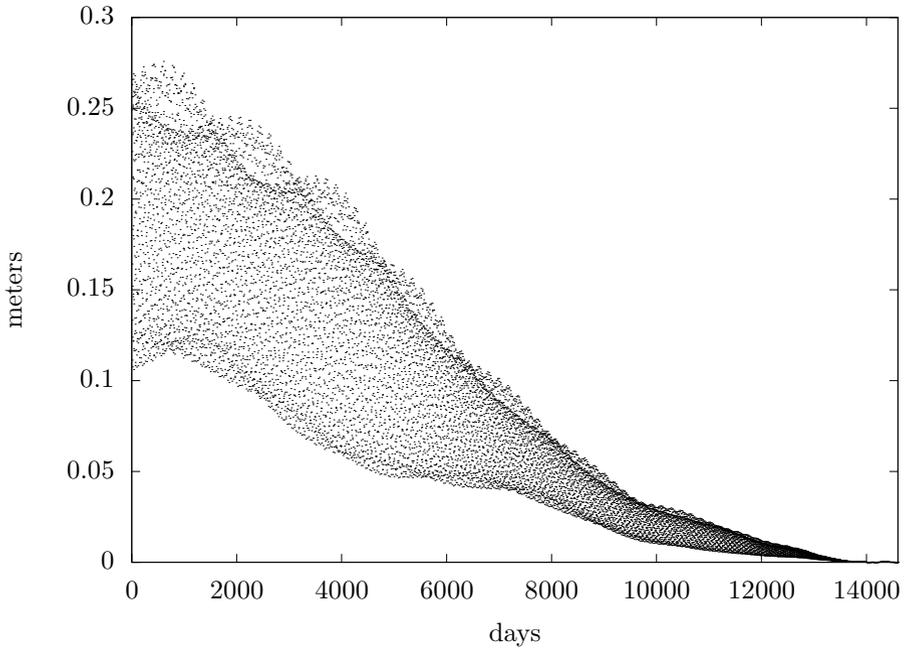}
  \caption{The roundoff errors of positions of Moon with double
    precision. Hereinafter, the position of the Moon is implied
    geocentric, while the positions of planets are barycentric.
  }\label{fig:diff-d-moon}
\end{figure}

\begin{figure}[p]
  \input{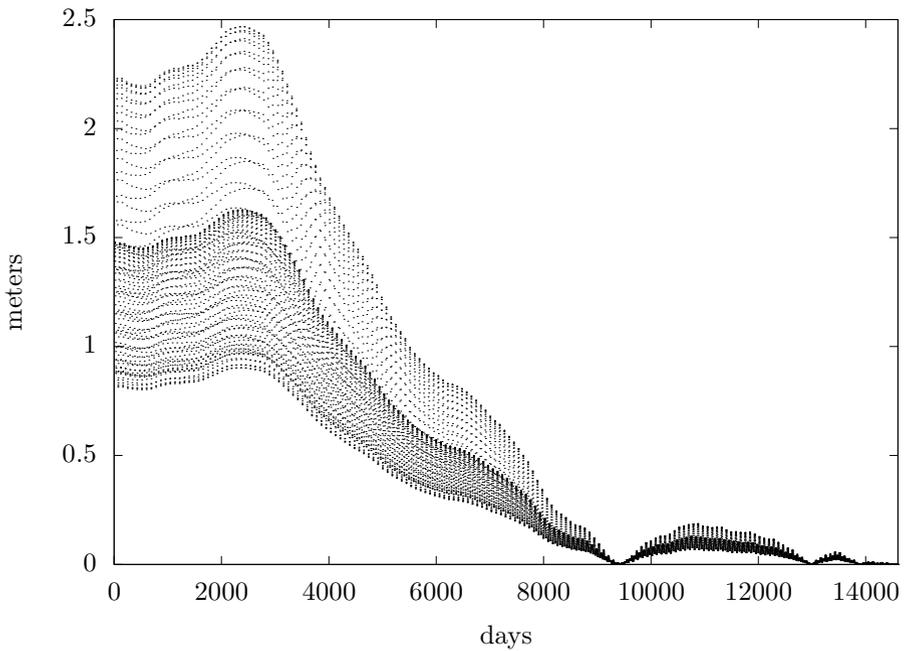}
  \caption{The roundoff errors of positions of Mercury with double precision}\label{fig:diff-d-mercury}
\end{figure}

\begin{figure}[p]
  \input{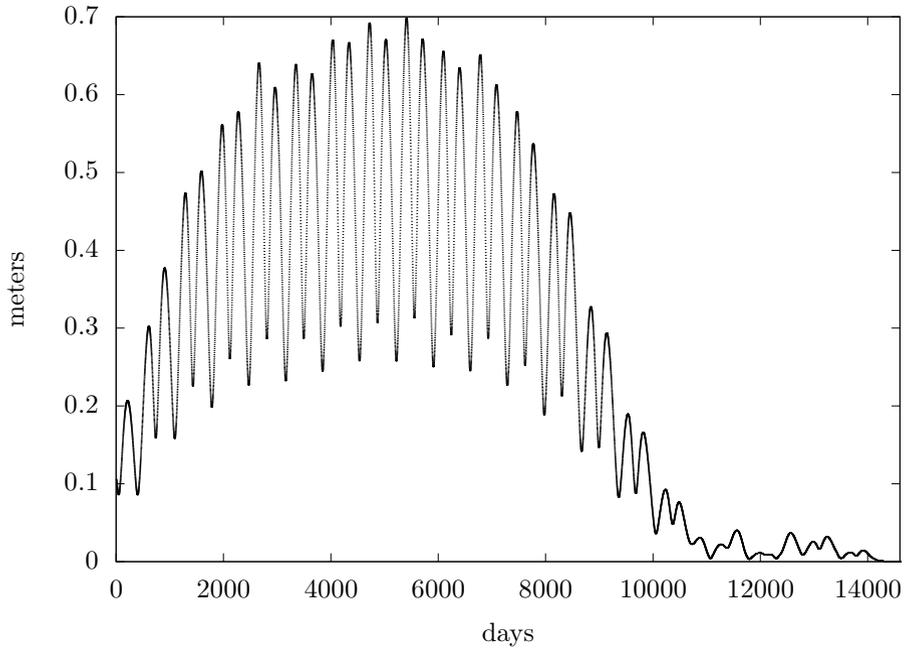}
  \caption{The roundoff errors of positions of Mars with double precision}\label{fig:diff-d-mars}
\end{figure}

The roundoff errors of positions of Moon, Mercury, and Mars obtained
with double precision and with the step size of $1/16$ days are shown on
Figs. \ref{fig:diff-d-moon}--\ref{fig:diff-d-mars}.  The one-way integration
for 40 years took 81 seconds in this configuration.
It is clear that the numerical roundoff errors from double precision
are not below the accuracy of observations for all the three bodies:
25 cm for the Moon (normal points accuracy reach 1 mm), 2.5 m for Mercury
(postfit rms of normal points reach 60 cm), and 70 cm for Mars
(postfit rms of normal points reach 55 cm).

\begin{figure}[p]
  \input{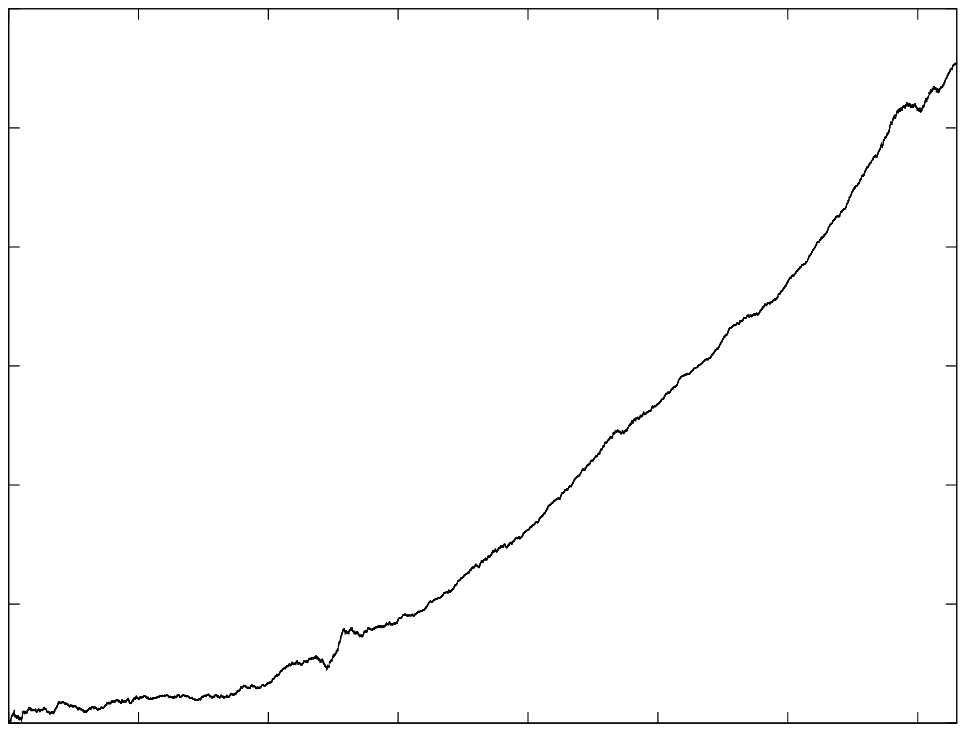}
  \caption{Drift of barycenter with double precision}\label{fig:barycenter-d}
\end{figure}

\subsection{Barycenter}
The (relativistic) barycenter of a system of N bodies is defined as follows:

\begin{equation}
  \bm{b} = \frac{\sum_{i} \mu_i^{*} \bm{r}_i}{\sum_{i}\mu_i^{*}}
\end{equation}

\noindent where $\mu_i^{*}$ is the body's relativistic gravitational parameter:

\begin{equation}
\mu_i^{*} = \mu_i\left(1+\frac{1}{2c^2}\dot{r_i}^2 -
\frac{1}{2c^2}\sum_{j\neq i}\frac{\mu_j}{r_\mathrm{ij}}\right)
\end{equation}

Initial positions and velocities of the bodies are equally adjusted
so that the position of the barycenter and the momentum are zero:

\begin{equation}
\begin{cases}
\sum_{i} \mu_i^{*} \bm{r}_i = \bm{0} \\
\sum_{i} \dot{\mu_i^{*}} \bm{r}_i + \mu_i^{*} \bm{\dot{r}}_i = \bm{0}
\end{cases}
\end{equation}

Mathematically, the barycenter of the system should stay in the origin
throughout the integration, but errors of approximation and numerical
roundoff cause it to drift. On Fig.~\ref{fig:barycenter-d}, we examine the distance from the
barycenter to the origin (as function of time) as an additional
consistency check of the correctness of the implementation of the EIH equations~(\ref{eq:eih})
  and the correctness and stability of the difference scheme~(\ref{eq:pred-corr}). The sub-millimeter
value of the drift confirms that the consistency check is passed.

\section{Reducing the roundoff errors}

\subsection{Existing approaches}
All mitigations of the problem of accumulating roundoff errors are
based, in one way or another, on using data types other than double
precision. DE ephemerides prior to DE405 (1997) were integrated on a
UNIVAC mainframe using its native 72-bit floating-point type with
60-bit fraction (coincidentally, that type was also called double
precision); DE405 was integrated on VAX Alpha using its native 128-bit
floating-point type with 112-bit fraction, \textit{quadruple
  precision} \cite{Chap8}. DE410 (2003) was integrated on Sun Ultra
with an UltraSPARC processor using ordinary double precision, which
caused the problems with numerical noise, limiting accuracy of orbits
at 1--2 meters~\cite{DE410}. The numerical integrator with which DE
was built was called DIVA~\cite{Krogh}. By the time of DE414 (2006),
DIVA was upgraded to QIVA, where ``Q'' stands for quadruple precision
(not implemented in hardware, but emulated by Fortran compiler). In
fact, a mixed mode was used: only the newtonian part of acceleration
was computed in quadruple precision, while all the remaining
perturbations were computed in double precision. Still, the mixed mode
required 30x more time than double precision mode. QIVA is probably
used to build DE ephemerides to this day, though compilers might since
have improved their emulation. It is important to note that quadruple
precision is very excessive for the task.

\begin{figure}[p]
  \input{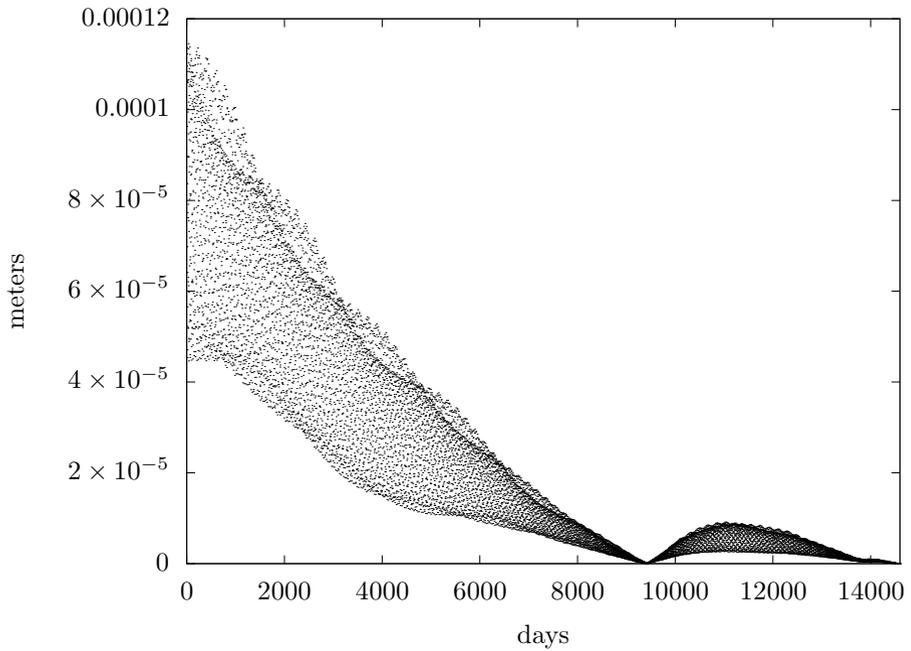}
  \caption{The roundoff errors of the position of the Moon with extended precision}\label{fig:diff-ld-moon}
\end{figure}

\begin{figure}[p]
  \input{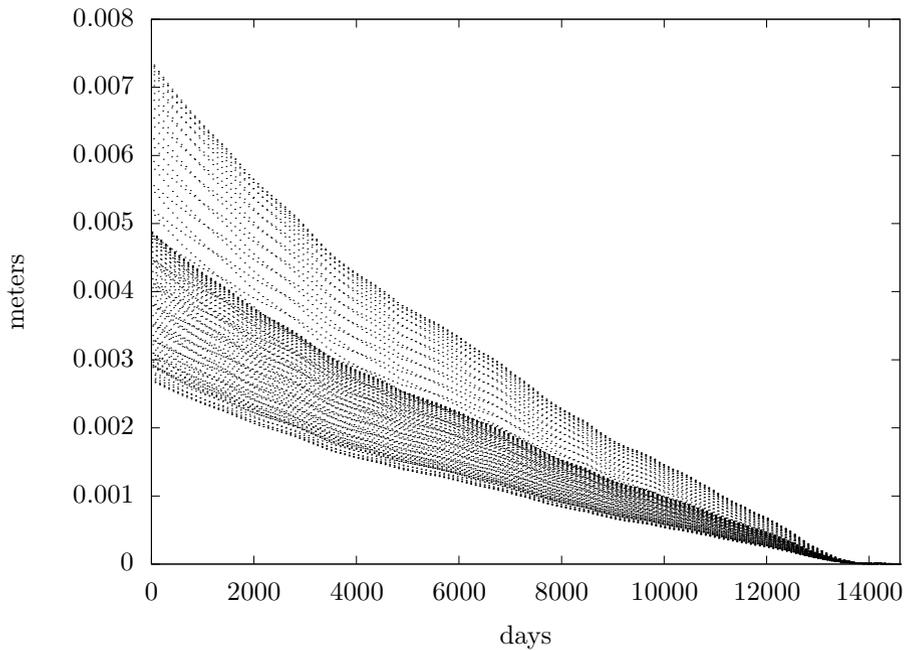}
  \caption{The roundoff errors of the position of Mercury with extended precision}\label{fig:diff-ld-mercury}
\end{figure}

\begin{figure}[p]
  \input{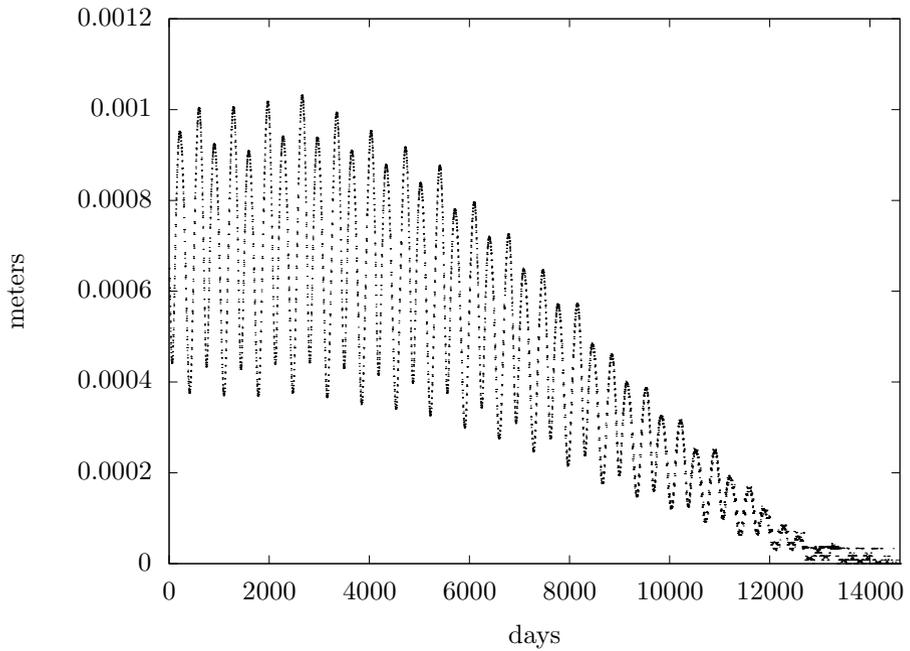}
  \caption{The roundoff errors of the position of Mars with extended precision}\label{fig:diff-ld-mars}
\end{figure}

\begin{figure}[p]
  \input{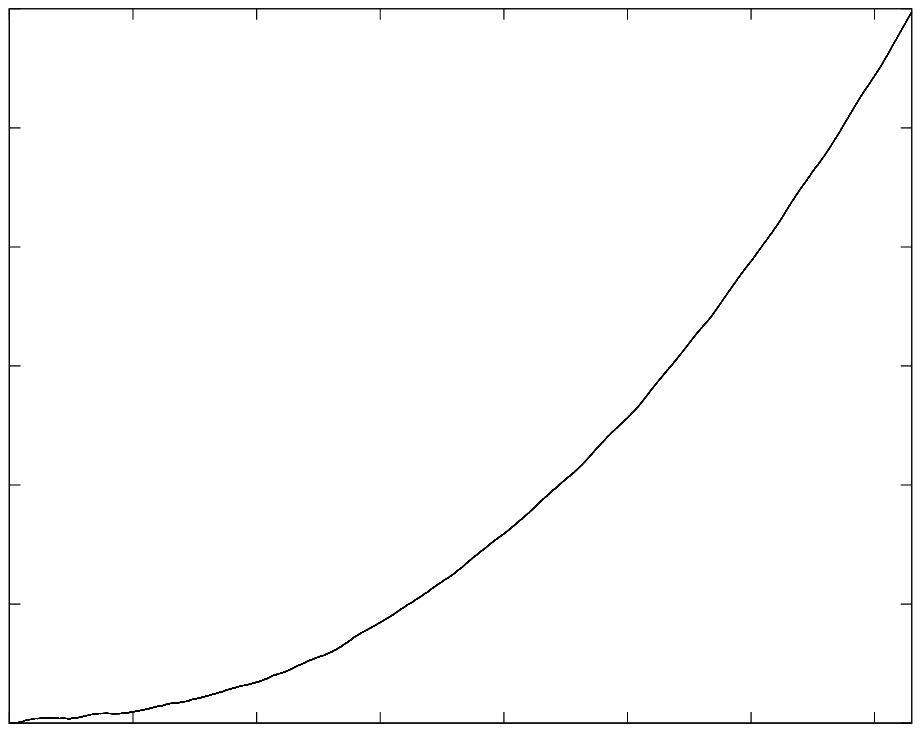}
  \caption{Drift of barycenter with extended precision}\label{fig:barycenter-ld}
\end{figure}

All public versions of EPM since EPM2004, as well as all versions of
INPOP since INPOP06, were numerically integrated with 80-bit
\textit{extended precision} type (63-bit fraction). This precision
is enough for planetary orbits, see Figs.~\ref{fig:diff-ld-moon}--\ref{fig:barycenter-ld}.
However, it has its own drawbacks:

\begin{itemize}
  \item Extended precision is supported exclusively on x86
    architecture (Intel and AMD processors), but not on ARM-based
    processors which seem to be gaining significant share (Apple M1,
    Microsoft SQ1, AWS Graviton, almost all mobile chips).
  \item Few programming languages provide support extended precision,
    namely Delphi, C/C++, D, Perl, Racket BC\footnote{``BC'' stands for ``before Chez'', the Racket compiler and virtual machine that existed before they were replaced with Chez Scheme, but are still supported.}, Swift, and some
    implementations of Common Lisp and Fortran. Extended precision in
    Fortran is limited to gfortran compiler; Intel and Absoft Fortran
    compilers do not support it.
  \item While simple operations with extended precision numbers are as
    fast as operations with double precision numbers in x86, SIMD
    extensions of x86 architecture (MMX, SSE, AVX, AVX-512) are not
    available for extended precision numbers. This limitation causes
    4x slowdown from double precision mode. (The integration in one
    direction for 40 years took 323 seconds.)

\end{itemize}

\subsection{Implementation of double-double arithmetic}

Double-double arithmetic is a software method to increase floating-point precision that uses pairs of double precision values to store numbers and performs arithmetic operation basing on arithmetics with double precision. Double-double numbers are defined as the unevaluated sums of two double components (larger
and smaller part): $x=x_{\mathrm{low}} + x_{\mathrm{high}}, \left \lvert x_{\mathrm{low}} \right \rvert \leq 0.5\cdot
\mathrm{ulp}(x_{\mathrm{high}})$ so their significands don't overlap. Such technique allows to store numbers with significands of at least $2 \cdot 53 = 106$ bits. 

Double-double arithmetic does not require special hardware nor programming language support
(software implementation \cite{QD} is available). The algorithms of some basic operations for the double-double type are shown below.

\begin{itemize}
    \item Dekker's summation or two double precision numbers with error \cite{float_book}. $\mathrm{RN}(x)$ means rounding to nearest.
    
        \begin{algorithm}[H]
            \caption{Fast2Sum ($a$, $b$), $a \geq b$}\label{alg:Fast2Sum}
            \begin{algorithmic}
                \State $s \gets RN(a + b)$
                \State $z \gets RN(s - a)$
                \State $t \gets RN(b - z)$
                \State return (s, t)
            \end{algorithmic}
        \end{algorithm}
    
    \item Dekker's double-double summation \cite{float_book}. The Fast2Sum algorithm is used to add up the large components and then to add the summation error to small components.
        \begin{algorithm}[H]
            \caption{DDSum $(x_h, x_l, y_h, y_l)$}\label{alg:DekkerAdd}
            \begin{algorithmic}
                \If{$\left \lvert x_h \right \rvert \ge \left \lvert y_h \right \rvert$}
                    \State $(r_h, r_l) \gets \mathrm{Fast2Sum}(x_h, y_h)$
                    \State $s \gets \mathrm{RN}(\mathrm{RN}(r_l + y_l) + x_l)$
                \Else
                    \State $(r_h, r_l) \gets \mathrm{Fast2Sum}(y_h, x_h)$
                    \State $s \gets \mathrm{RN}(\mathrm{RN}(r_l + x_l) + y_l)$
                \EndIf
                \State return $\mathrm{Fast2Sum}(r_h, s)$
            \end{algorithmic}
        \end{algorithm}
   
    \item Double multiplication with error (uses \textit{fused multiply-add} instruction, or FMA).
        \begin{algorithm}[H]
            \caption{Product($x$, $y$)}\label{alg:Product}
            \begin{algorithmic}
                \State $p \gets x \cdot y$
                \State $e \gets \mathrm{FMA}(x, y, -p)$
                \State return $(p, e)$
            \end{algorithmic}
        \end{algorithm}
        
    \item Double-double multiplication \cite{float_book}.
        \begin{algorithm}[H]
            \caption{DDProduct $(x_h, x_l, y_h, y_l)$}
            \begin{algorithmic}
                \State $(c_h, c_l) \gets \mathrm{Product}(x_h, y_h)$
                \State $p_1 \gets \mathrm{RN}(x_h \cdot y_l)$
                \State $p_2 \gets \mathrm{RN}(x_l \cdot y_h)$
                \State $c_l \gets \mathrm{RN}(c_l + RN(p_1 + p_2))$
                \State return $\mathrm{Fast2Sum}(c_h, c_l)$
            \end{algorithmic}
        \end{algorithm}    
    
    \item Double-double division \cite{QD}. The first approximation is $q_0 = x_h / y_h$.
      The remainder is $r = x - q_0 \cdot y$. The correction term is $q_1 = r_h / b_h$.
      Finally, $\mathrm{Fast2Sum}(q_0, q_1)$ renormalizes the result.
      TwoOneProd is the product of double-double and double (written similarly to DDProduct).
      2Sum is the summation for
      cases where order of arguments for Fast2Sum is unknown.
      
        \begin{algorithm}[H]
            \caption{DDDiv $(x_h, x_l, y_h, y_l)$}
            \begin{algorithmic}
                \State $t_h \gets \mathrm{RN}(x_h / y_h)$
                \State $(r_h, r_l) \gets \mathrm{TwoOneProd}(y_h, y_l, t_h)$
                \State $(\pi_h, \pi_l) \gets \mathrm{2Sum}(x_h, -r_h)$
                \State $\sigma_h \gets \mathrm{RN}(\pi_l - r_l)$
                \State $\sigma_l \gets \mathrm{RN}(\sigma_h + x_l)$
                \State $\sigma \gets \mathrm{RN}(\pi_h + \sigma_l)$
                \State $t_l \gets \mathrm{RN}(\sigma/y_h)$
                \State return $\mathrm{Fast2Sum}(t_h, t_l)$
            \end{algorithmic}
        \end{algorithm}
    \item Double-double square root. The QD library~\cite{QD} uses
      Karp's trick for square root~\cite{KarpSQRT}.        
        If $s$ is approximation for $\sqrt{x}$ (QD uses $\sqrt{x_h}$), then $\sqrt{x} \approx x \cdot s + [x - (x\cdot s)^2] \cdot s / 2$. 
\end{itemize}

\begin{figure}[p]
  \input{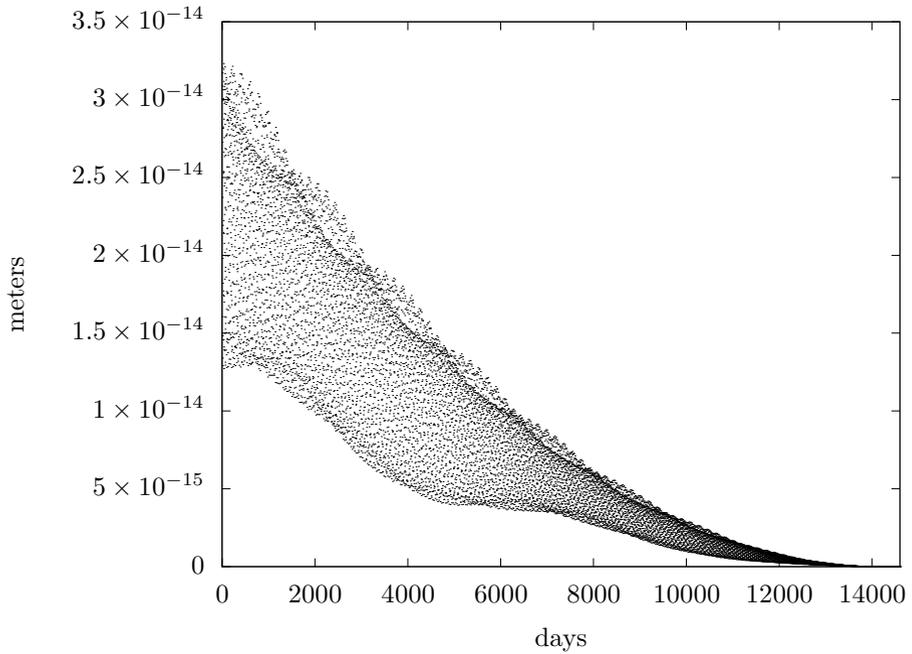}
  \caption{The roundoff errors of the position of the Moon with double-double precision}\label{fig:diff-dd-moon}
\end{figure}

\begin{figure}[p]
  \input{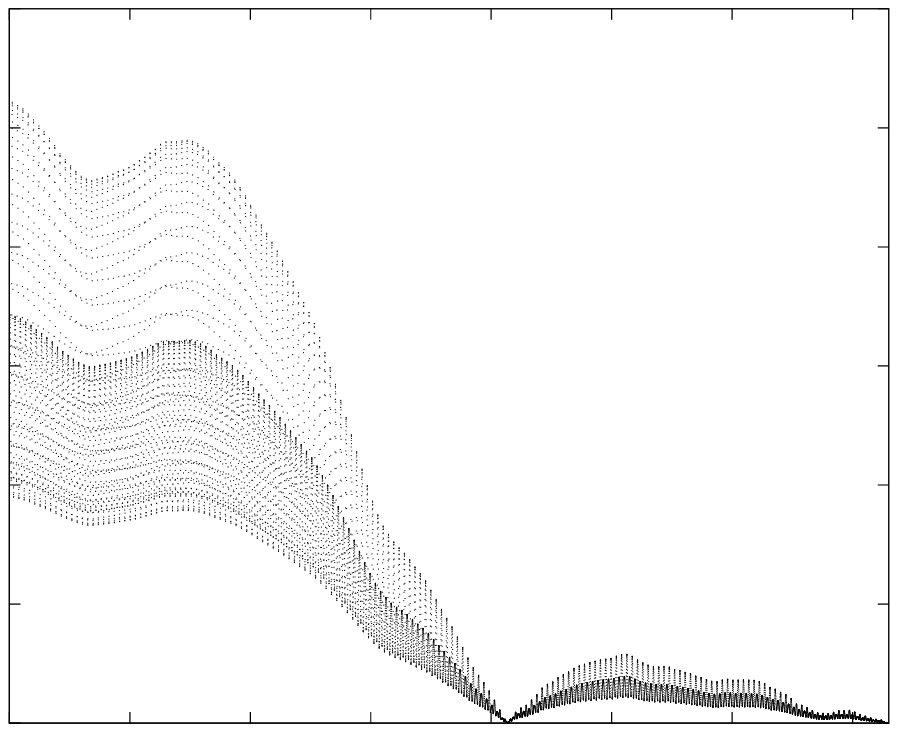}
  \caption{The roundoff errors of the position of Mercury with double-double precision}\label{fig:diffdd-mercury}
\end{figure}

\begin{figure}[p]
  \input{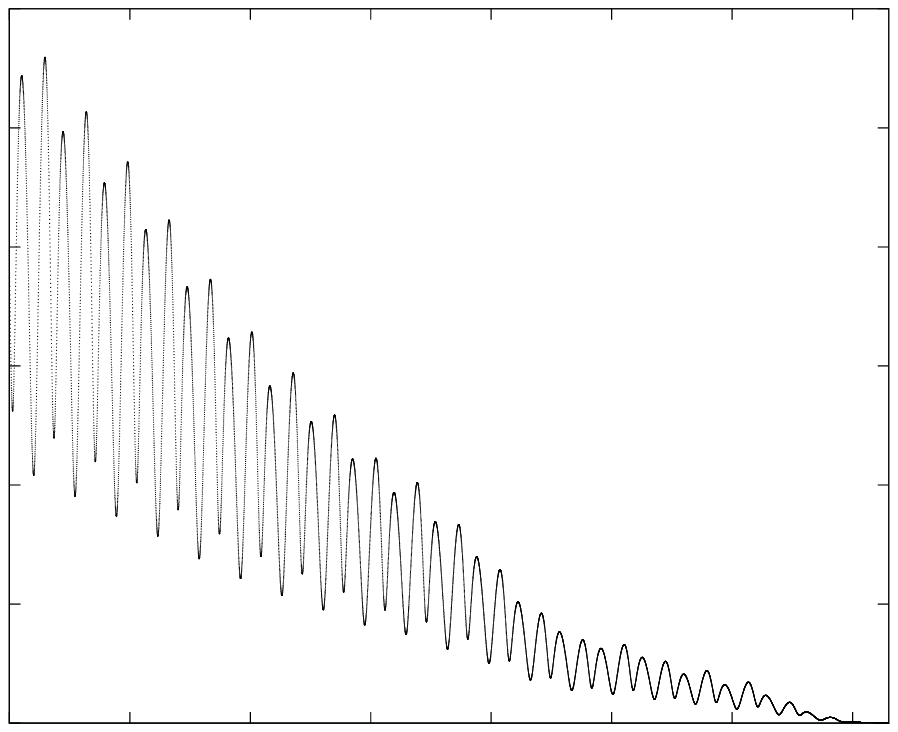}
  \caption{The roundoff errors of the position of Mars with double-double precision}\label{fig:diff-dd-mars}
\end{figure}

\begin{figure}[p]
  \input{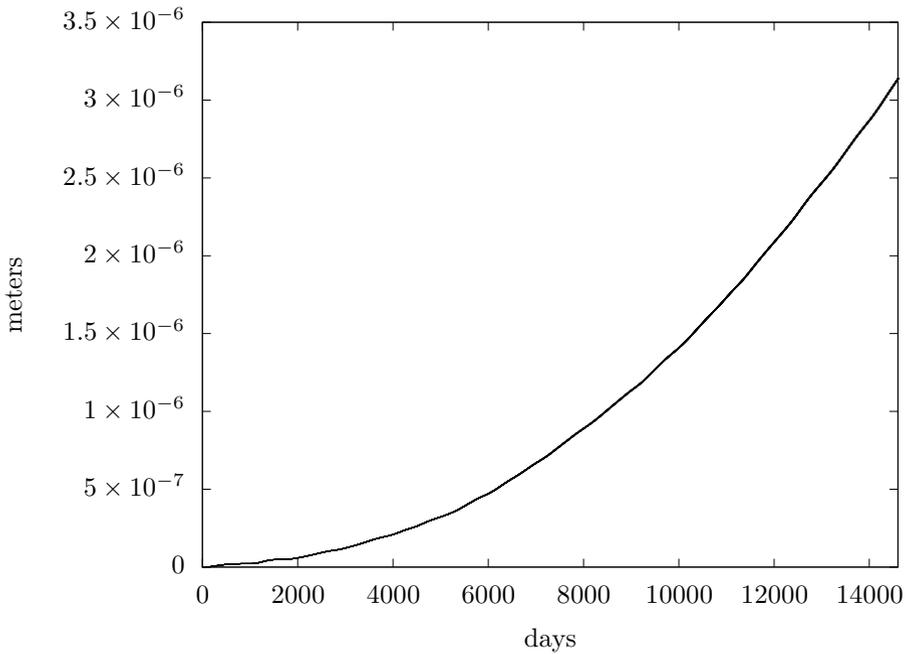}
  \caption{Drift of barycenter with double-double precision}\label{fig:barycenter-dd}
\end{figure}

As part of the experiment, a full double-double implementation of the
numerical integrator was done in this work. The results can be seen on
Figs.~\ref{fig:diff-dd-moon}--\ref{fig:barycenter-dd}. Clearly, the
numerical roundoff errors are negligible, but this comes at great
performance cost: the one-way integration for 40 years took 32
minutes, which is a 24x slowdown from double precision.

\subsection{Selective usage of double-double types}\label{sec:selective}
By trial and error, the following scheme was worked out to maintain the balance between
the precision and performance of calculations:
\begin{itemize}
  \item The state of the system $\bm{x}$ and its time derivative
    $\bm{f}$ are stored in double-double precision, as well as the
    intermediate values in the PECEC scheme.
\item The finite differences $\nabla^i$ are stored in double
  precision. However, the coefficients $\gamma_i$ are stored in
  double-double precision.
\item In the calculation of $\bm{f}(\bm{x})$, i.e. in Eq.~(\ref{eq:eih}),
  the $\bm{r}_{ij}$ vectors and $r_{ij}$ distances are stored in double precision.
  %(except
  %for the Earth--Moon distance, see Sec.~\ref{sec:earthmoon}).
\item In the relativistic part of Eq.~(\ref{eq:eih}), $\bm{v}_i$ and $\bm{a}_i$ are
  rounded down to double precision, and all calculations are performed in double precision.
\item The constant $1/c^2$ and the gravitational parameters $Gm_i$ are
  also stored in double precision.
\end{itemize}

\subsection{Special treatment of the Earth--Moon system}\label{sec:earthmoon}
Let us denote $\bm{r}_{\mathrm{EM}} = \bm{r}_{\mathrm{M}} -
\bm{r}_{\mathrm{E}}$ the Earth--Moon vector. All $\bm{r}_i$ in
Eq.~(\ref{eq:eih}) are supposed to be referred to a single inertial
frame (barycentric celestial reference system, or BCRS, in our
case). However, a direct calculation of $\bm{r}_{\mathrm{M}} -
\bm{r}_{\mathrm{E}}$ will result in a value that is about 400 times
smaller than the subtrahend and the minuend, thus causing the loss of
precision from the ``catastrophic cancellation'' of significant bits.
This is particularly unwanted for $\bm{r}_{\mathrm{EM}}$ because of
the high requirements to the numerical accuracy of the Moon's orbit.
In all three families of ephemerides (DE, EPM, INPOP) the orbit of the
Moon is integrated in geocenter, so that $\bm{r}_{\mathrm{EM}}$ enters
$\bm{x}$ while the barycentric $\bm{r}_{\mathrm{M}}$ is temporarily
restored during the calculation of $\bm{f}$.

Similarly, we aim to avoid the catastrophic cancellation in the calculation
of the Newtonian acceleration of the geocentric Moon. For mutual Newtonian
forces between Earth and Moon, this is rather simple. In Eq.~(\ref{eq:aem}),
%both $\bm{a}_{\mathrm{EM}}$ and $\bm{r}_{\mathrm{EM}}$ are stored in double-double precision and
no subtraction of close numbers is involved.

\begin{equation}\label{eq:aem}
  {\bm{a}}_{\mathrm{EM}}^{\mathrm{(E,M)}} = {\bm{a}}_\mathrm{M}^{\mathrm{(E)}} - {\bm{a}}_\mathrm{E}^{\mathrm{(M)}} = -\frac{Gm_\mathrm{M} + Gm_\mathrm{E}}{r_{\mathrm{EM}}^3}\bm{r}_{\mathrm{EM}}
  \end{equation}

For geocentric Moon acceleration exerted by other bodies, we can apply
a Taylor expansion. For body A, we aim to avoid direct subtraction of
the Newtonian accelerations of the Moon
$\bm{a}_\mathrm{M}^{\mathrm{(A)}}$ and Earth
$\bm{a}_\mathrm{E}^{\mathrm{(A)}}$ caused by this body:

\begin{equation}
  \begin{aligned}
    \bm{a}_{\mathrm{EM}}^{\mathrm{(A)}} &= {\bm{a}}_\mathrm{M}^{\mathrm{(A)}} - {\bm{a}}_\mathrm{E}^{\mathrm{(A)}} 
  =    Gm_\mathrm{A}\left(\frac{\bm{r}_\mathrm{MA}}{{r}_{\mathrm{MA}}^3} - \frac{\bm{r}_\mathrm{EA}}{r_{\mathrm{EA}}^3} \right) \\
  &= \frac{Gm_\mathrm{A}}{r_{\mathrm{MA}}^3}\left[\bm{r}_\mathrm{MA} - \bm{r}_\mathrm{EA}\left(\frac{(\bm{r}_\mathrm{EA} - \bm{r}_\mathrm{EM})^2}{r_{\mathrm{EA}}^2}\right)^{3/2}\right]\\
  &= \frac{Gm_\mathrm{A}}{r_{\mathrm{MA}}^3}\Bigg[\bm{r}_\mathrm{MA} - \bm{r}_\mathrm{EA}\bigg(1 - 2\frac{\bm{r}_\mathrm{EA}\bm{r}_\mathrm{EM}}{r_{\mathrm{EA}}^2} +  \frac{r_\mathrm{EM}^2}{r_\mathrm{EA}^2} \bigg)^{3/2}\Bigg]
  \end{aligned}
\end{equation}

Denoting $x = 2\frac{\bm{r}_\mathrm{EA}\bm{r}_\mathrm{EM}}{r_{\mathrm{EA}}^2} -  \frac{r_\mathrm{EM}^2}{r_\mathrm{EA}^2}$ and applying the Taylor expansion of $(1-x)^{3/2}$, we get

\begin{equation}\label{eq:aem-final}
  \bm{a}_{\mathrm{EM}}^{\mathrm{(A)}} \approx \frac{Gm_\mathrm{A}}{r_{\mathrm{MA}}^3}\left[ - \mathbf{r}_\mathrm{EM} - \mathbf{r}_\mathrm{EA} \left(-\frac{3x}{2} + \frac{3x^2}{8} + \frac{x^3}{16} + \frac{3x^4}{128}\right)\right]
  \end{equation}
\noindent in which there is no subtraction of close values. Also, it
was found out that double precision is enough for
Eq.~(\ref{eq:aem-final}) and that four terms of Taylor expansion are
required, but not more.

\subsection{Results}\label{sec:results}
\begin{figure}[p]
  \input{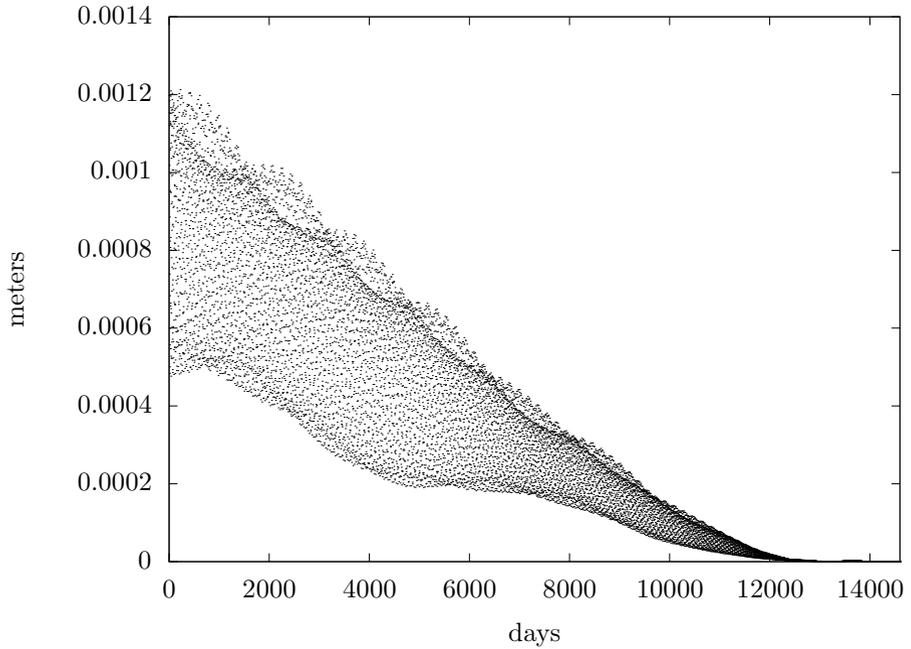}
  \caption{The roundoff errors of the position of the Moon with mixed precision}\label{fig:diff-mixed-moon}
\end{figure}

\begin{figure}[p]
  \input{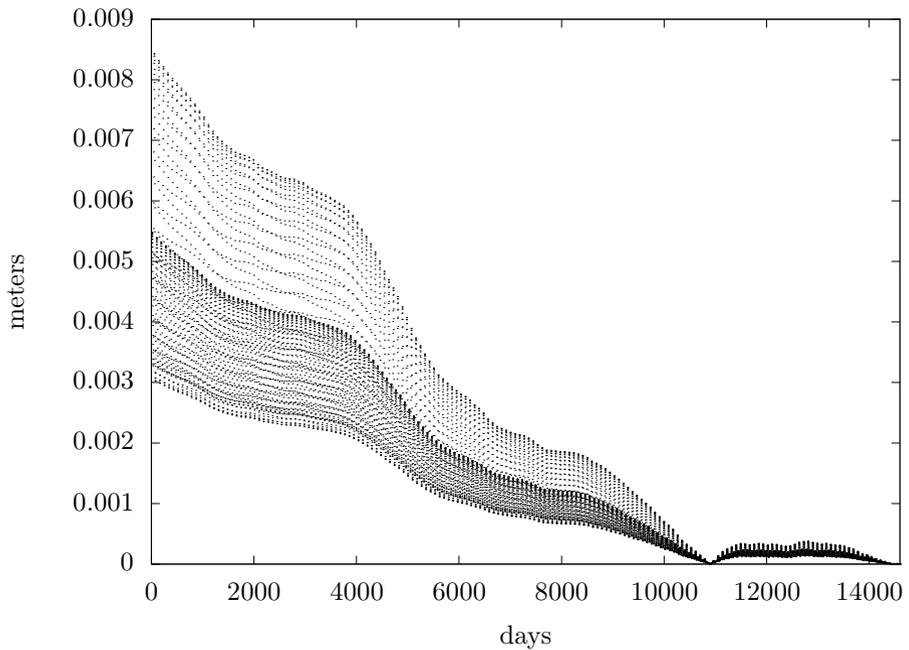}
  \caption{The roundoff errors of the position of Mercury with mixed precision}\label{fig:diff-mixed-mercury}
\end{figure}

\begin{figure}[p]
  \input{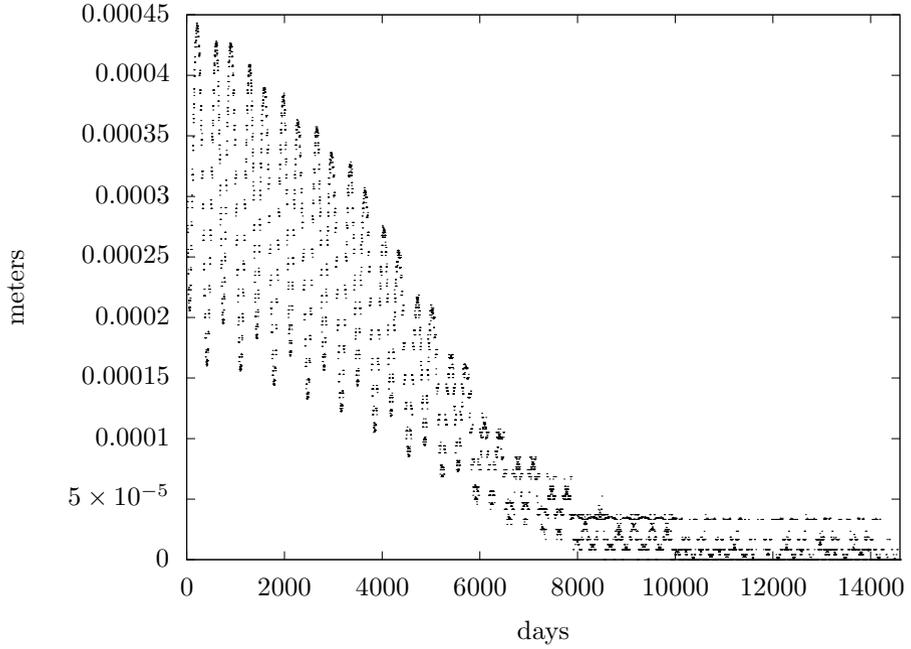}
  \caption{The roundoff errors of the position of Mars with mixed precision}\label{fig:diff-mixed-mars}
\end{figure}

\begin{figure}[p]
  \input{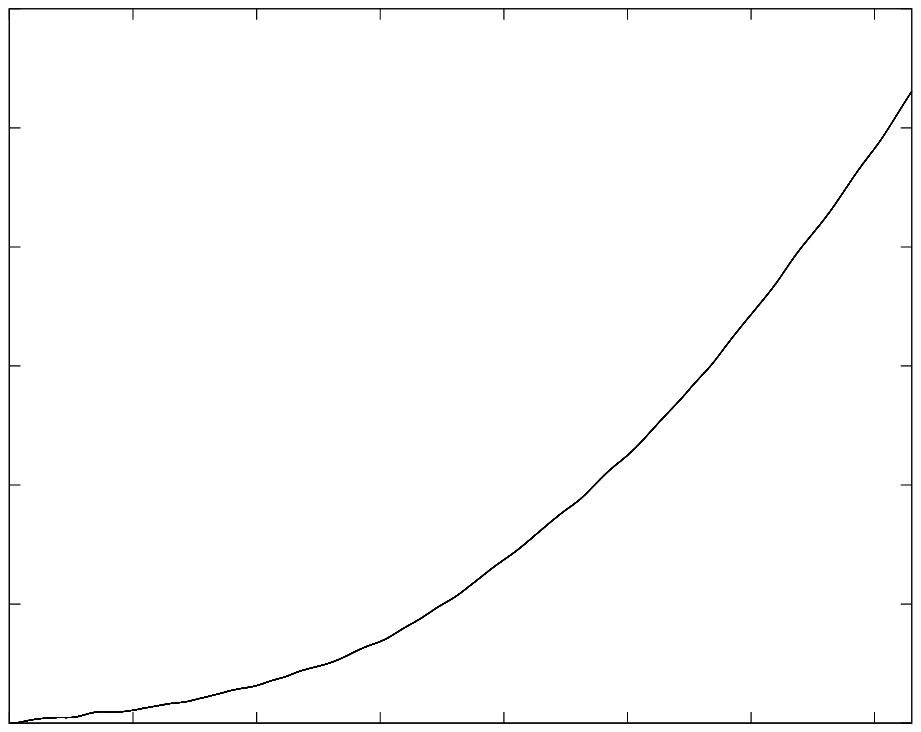}
  \caption{Drift of barycenter with mixed precision}\label{fig:barycenter-mixed}
\end{figure}

The final results obtained with the mixed precision approach (Sec.~\ref{sec:selective})
and with the special treatment of the Earth--Moon system (Sec.~\ref{sec:earthmoon})
are shown of Figs.~\ref{fig:diff-mixed-moon}--\ref{fig:barycenter-mixed}.
The accuracy is comparable to that obtained with extended precision. The one-way
integration for 40 years took 246 seconds.

\section{Conclusion}
We summarize the results in Table~\ref{tbl:results}. It can be
clearly seen that the integration in ``mixed'' mode developed
in this work gives the numerical roundoff errors that are small enough
for the modern ephemeris astronomy, while not requiring
any primitive datatypes other than the 64-bit double precision.

The ``mixed'' mode is just 3 times slower than the double precision
mode and 1.3 times faster than the extended precision
mode. The latter can be explained by the fact that the extended
precision does not enjoy SIMD extensions in x86 hardware.

It must be noted that the results of such measurements are generally
affected by the equations used (the number of asteroids and other
point masses, figure effects, etc). The figure effects are
particularly important for the orbital and rotational motion of the
Moon. Also, derivatives of $\bm{x}$ w.r.t. model parameters are often
integrated alongside the dynamical equations. All this additional
complexity was omitted in this work; however, it is not likely to
change the results in principle. Equations of the rotation of the Moon
can be easily implemented in double-double without significant impact
on overall performance; figure effects in acceleration are small
enough to be computed in double precision; and for the derivatives,
high precision is not required.

\begin{table}
\begin{tabular}{lllll}
 Precision & \textbf{double} & \textbf{extended} & \textbf{mixed}  & \textbf{double-double} \\
 Max. error for Moon & 0.28 m & 0.1 mm & 1.2 mm & $3.2 \cdot 10^{-14}$ m\\
 Max. error for Mercury & 2.47 m & 7 mm & 8.4 mm & $2.6 \cdot 10^{-16}$ m\\
 Max. error for Mars & 0.7 m & 1 mm & 0.44 mm & $2.8 \cdot 10^{-16}$ m\\ 
 Max. drift of barycenter & 0.5 mm & 2.9 $\mu$m & 2.6 $\mu$m & 3.14 $\mu$m\\
 Time of integration & 81 s & 323 s & 246 s & 32 min
\end{tabular}
\caption{Comparative results for 40 years of numerical integration
  with different data types.} \label{tbl:results}
  \end{table}

\backmatter

\section*{Data availability}
The program code and data used for numerical integration in this work are available
online: \url{https://github.com/S1ckick/Nbodies}.

%\bmhead{Acknowledgments}

%\section*{Declarations}

%% Some journals require declarations to be submitted in a standardised
%% format. Please check the Instructions for Authors of the journal to
%% which you are submitting to see if you need to complete this
%% section. If yes, your manuscript must contain the following sections
%% under the heading `Declarations':

%% \begin{itemize}
%% \item Funding
%% \item Conflict of interest/Competing interests (check journal-specific guidelines for which heading to use)
%% \item Ethics approval 
%% \item Consent to participate
%% \item Consent for publication
%% \item Availability of data and materials
%% \item Code availability 
%% \item Authors' contributions
%% \end{itemize}

\bibliography{references}
\end{document}